\documentclass[conference]{IEEEtran}
\IEEEoverridecommandlockouts
\usepackage{cite}
\usepackage{amsmath,amssymb,amsfonts}
\usepackage{algorithmic}
\usepackage{graphicx}
\usepackage{textcomp}
\usepackage{xcolor}
\usepackage{bbm}
\usepackage{bm}
\usepackage{subfigure}
\usepackage{pifont}
\usepackage{algorithm} 

\usepackage[switch]{lineno}

\def\BibTeX{{\rm B\kern-.05em{\sc i\kern-.025em b}\kern-.08em
    T\kern-.1667em\lower.7ex\hbox{E}\kern-.125emX}}

\begin{document}

\title{VDTuner: Automated Performance Tuning for Vector Data Management Systems\\
}

\author{
Tiannuo Yang\textsuperscript{1}, Wen Hu\textsuperscript{2}, Wangqi Peng\textsuperscript{1}, Yusen Li\textsuperscript{1}, Jianguo Li\textsuperscript{2}, Gang Wang\textsuperscript{1}, and Xiaoguang Liu\textsuperscript{1} \\
\textsuperscript{1} \textit{College of Computer Science, TMCC, SysNet, DISSec, GTIISC, Nankai University, Tianjin, China} \\
\textsuperscript{2} \textit{Ant Group, China} \\
Email: \{yangtn, pengwq, liyusen, wgzwp, liuxg\}@nbjl.nankai.edu.cn, \{huwen.hu, lijg.zero\}@antgroup.com

\thanks{This work was done when Tiannuo Yang and Wangqi Peng were research interns at Ant Group. Yusen Li and Jianguo Li are the corresponding authors.}
}


\maketitle

\begin{abstract}
Vector data management systems (VDMSs) have become an indispensable cornerstone in large-scale information retrieval and machine learning systems like large language models. To enhance the efficiency and flexibility of similarity search, VDMS exposes many tunable index parameters and system parameters for users to specify. However, due to the inherent characteristics of VDMS, automatic performance tuning for VDMS faces several critical challenges, which cannot be well addressed by the existing auto-tuning methods. 

In this paper, we introduce VDTuner, a learning-based automatic performance tuning framework for VDMS, leveraging multi-objective Bayesian optimization. VDTuner overcomes the challenges associated with VDMS by efficiently exploring a complex multi-dimensional parameter space without requiring any prior knowledge. Moreover, it is able to achieve a good balance between search speed and recall rate, delivering an optimal configuration. Extensive evaluations demonstrate that VDTuner can markedly improve VDMS performance (14.12\% in search speed and 186.38\% in recall rate) compared with default setting, and is more efficient compared with state-of-the-art baselines (up to 3.57$\times$ faster in terms of tuning time). In addition, VDTuner is scalable to specific user preference and cost-aware optimization objective. VDTuner is available online at https://github.com/tiannuo-yang/VDTuner.
\end{abstract}


\section{Introduction}

In recent years, the emergence of large language models (LLM) has brought the development of AI technology to an unprecedentedly thriving stage \cite{openai2023gpt4,zhao2023llmsurvey}.
In the LLM application scenarios, unstructured multimedia data is usually converted into embedding vectors as a strong knowledge base to overcome the hallucination problem in conversations \cite{huang2023hallucinationsurvey,cui2023chatlaw,han2023vdbsurvey,bang2023evaluatechatgpt}. Consequently, various purpose-built vector data management systems (VDMSs) \cite{wang2021milvus, guo2022manu, qdrant2023, chroma2023} have been developed to provide efficient, scalable, and reliable management of these vectors. The booming development of LLM applications makes the high efficient VDMS become the infrastructure of the LLM-era. Many popular VDMSs, such as Milvus \cite{wang2021milvus} and Qdrant \cite{qdrant2023}, now possess a substantial user base with an especially active community. 

VDMS has three distinguished features. First, like many traditional databases, VDMS usually exposes many tunable system parameters which have significant impacts on performance. Second, particularly built for similarity search with massive vector data, VDMS involves an important indexed query step, which requires users to specify one index type and several index parameters. Third, VDMS contains two critical performance metrics simultaneously: search speed and recall rate. Although the auto-configuration of traditional databases have been extensively studied \cite{van2017ottertune, zhang2019cdbtune, trummer2022dbbert, zhang2023unitune, pgtune2023, li2019qtune, cereda2021cgptuner, zhao2022dremel, guo2022manu}, none of the existing literature considers dedicated auto-configuration solution for VDMS. Thus, in this paper, we aim to address the problem: how can VDMS parameters be automatically configured to maximize both search speed and recall rate?

While we observe that auto-configuring VDMS has great performance improvement potential, designing an optimally efficient performance tuning approach for VDMS is non-trivial. First, the parameters of VDMS are intricately interdependent, thus finding the optimal VDMS configuration requires exploring complex multi-dimensional search space. Second, VDMS has two conflicting performance metrics (i.e., recall rate and search speed), and simultaneously optimizing the two metrics is challenging. Third, tunable parameters are different for different index types. To identify the most suitable index type within limited tuning budgets is challenging.

Many auto-tuning solutions have been proposed for improving the performance of traditional databases \cite{zhao2023dbtuningsurvey}. However, they all fall short on efficiency and optimality for tuning VDMS. The naive-search methods such as random and simulated annealing lack efficiency because they can not use historical information effectively. 
Although heuristic strategies \cite{ansel2014opentuner, zhu2017bestconfig}, which usually employ some numerical optimization techniques, incur very low overhead, they commonly suffer from unstable performance and local optimum as they failed to consider the complex dependency between parameters.
Learning strategies such as Bayesian optimization \cite{van2017ottertune, cereda2021cgptuner, zhang2021restune} and reinforcement learning \cite{zhang2019cdbtune, li2019qtune} have also gained a lot of attention in database tuning. 
Although these leaning strategies can learn complex configuration space to achieve superior performance, they still lack efficiency when facing the conflicting objectives and non-fixed parameter space for different index types in VDMS tuning.



To address the above challenges, we propose VDTuner, an auto-tuning framework for VDMS, aiming to maximize both search speed and recall rate.  VDTuner has many promising features which are highly suitable for VDMS tuning: (1) it does not require prior knowledge about VDMS; (2) it is able to explore complex multi-dimensional  parameter space efficiently; (3) it is able to strike a good balance between search speed and recall rate.  The key idea of VDTuner is to leverage Multi-objective Bayesian Optimization (MOBO) \cite{yang2019mobo, daulton2020qehvi}, which is a popular technique widely used for  solving multi-objective optimization problem with expensive and black-box function. However, applying MOBO to VDMS tuning is not easy (section \ref{sec:mobo_challenge}), and we have proposed many new techniques to address these challenges (section \ref{sec:vdtuner_desgin}).

We conduct extensive experiments to evaluate VDTuner. Results prove that VDTuner can significantly improve search speed and recall rate (up to 14.12\% and 186.38\%, respectively) compared with default setting, which confirms the necessity of auto-configuring VDMS. Moreover, VDTuner outperforms state-of-the-art baselines in terms of VDMS performance and tuning efficiency with a significant margin (1.48$\times$ to 3.57$\times$ faster). 

In summary, we make the following major contributions:
\begin{itemize}
    \item We perform extensive preliminary studies to identify the main challenges of VDMS tuning and analyze the deficiencies of existing solutions in VDMS tuning.
    \item We propose VDTuner, a performance tuning framework for VDMS, which utilizes Multi-objective Bayesian Optimization to discover the optimal configuration of VDMS, maximizing both search speed and recall rate.
    \item We comprehensively evaluate VDTuner to verify its performance and analyze the reason of its efficiency. We demonstrate that VDTuner beats all baselines with a significant margin.
\end{itemize}

\section{Background and Motivation}
\subsection{Vector Data Management System (VDMS)}
Vector database management system (VDMS) is a purpose-built data management system to efficiently manage large-scale vector data \cite{wang2021milvus}. The most common operation in VDMS is similarity search, i.e., searching for the top $K$ similar vectors from stored data given a new vector. VDMS is featured mainly by the following aspects.

\textbf{Multiple Components.} To enhance elasticity and flexibility, modern VDMS architecture usually consists of multiple layers (e.g., Access, Coordinator, Worker and Storage) dedicated for specific functions. Each layer has many components working together. 
For example, within Milvus, the data coordinator and index coordinator components manage the topology of data nodes and index nodes, respectively.
These components expose many tunable \textbf{\textit{system parameters}} of VDMS for users to specify, as a fixed configuration may not be applicable to all scenarios. 

\textbf{Multiple Index Types.} Similarity search is noted for its high complexity. Except for brute-force search, many approximate nearest neighbor search (ANNS) algorithms (e.g., Product Quantization) are integrated into VDMS to improve the performance of similarity search. Each ANNS algorithm requires its own index type, hence, VDMS usually need to maintain multiple index types. For instance, the index type IVF\_PQ internally employs a Product Quantization algorithm, while HNSW utilizes a graph-based ANNS algorithm known as Hierarchical Navigable Small World Graph.

A complete indexed query process requires users to specify one \textbf{\textit{index type}} (e.g., HNSW) and several \textbf{\textit{index parameters}} (e.g., node degree \textit{M} and search scope \textit{efConstruction}).
In this manner, VDMS dramatically accelerates the time-consuming similarity search on large datasets. 

    
\textbf{Multiple Performance Metrics.} Different ANNS algorithms present different search speed (i.e., the request number that VDMS can handle per second) and recall rate (i.e., the ratio of correctly retrieved similar vectors to the total actual similar vectors). In addition, different parameters of a specific ANNS algorithm also results in different search speed and recall rate. Therefore, VDMS users usually care about two metrics simultaneously: search speed and recall rate.

 
\subsection{Challenges in Auto-Configuring VDMS}
While the parameter setting of VDMS can greatly impact its performance, auto-configuring VDMS faces many challenges. 

\vspace{4pt}
\textbf{Challenge 1.} \textit{VDMS parameters are intricately inter-dependent, thus finding the optimal VDMS configuration requires exploring a complex and multi-dimensional search space.}
\vspace{4pt}

Take the popular VDMS, Milvus \cite{wang2021milvus}, as an example, the index and system parameters recommended for performance tuning have totally 16 dimensions, and the values of most parameters are continuous. The resulting space is prohibitively large, thus exhausting all possible configurations becomes impossible. An alternative approach is to consider each parameter separately to reduce searching complexity. Unfortunately, this is infeasible since VDMS configurations are intricately interdependent with each other. 

\begin{figure} 
\centerline{\includegraphics[width=.48\textwidth]{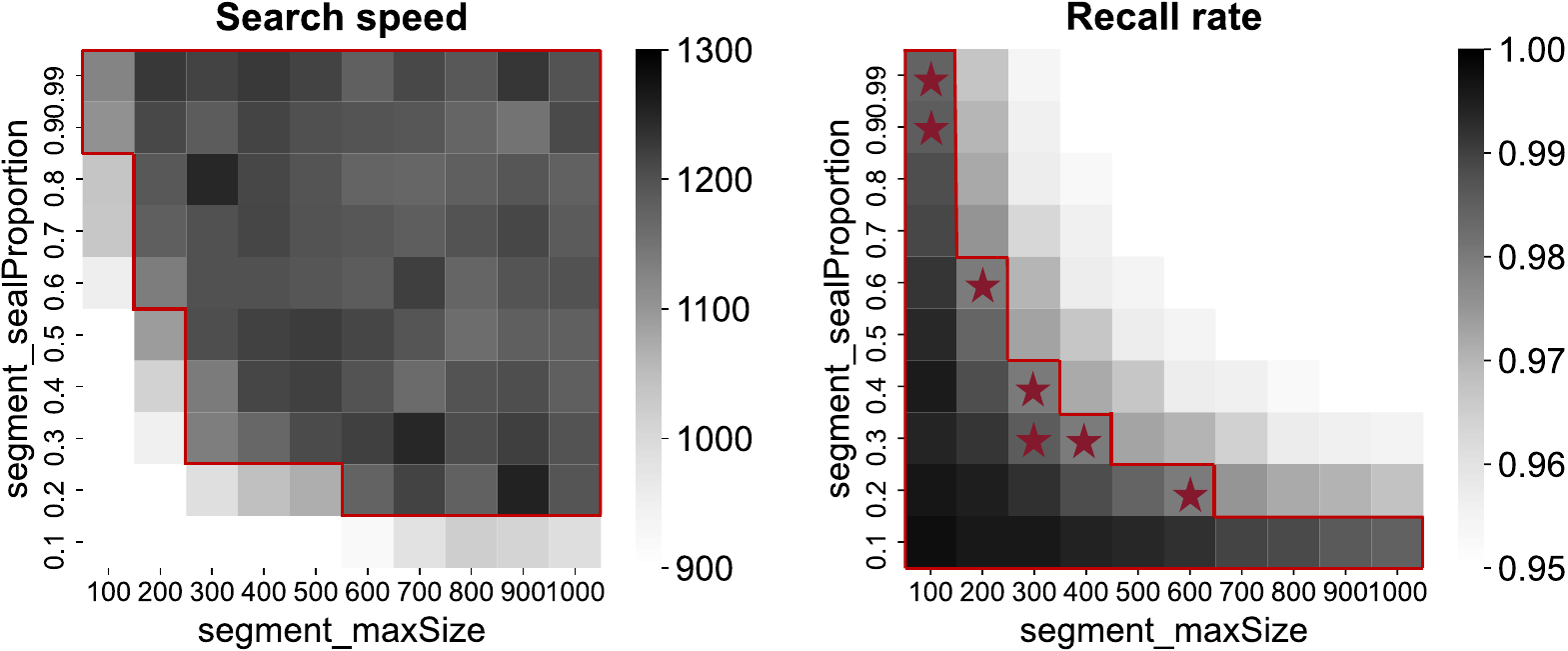}}
\caption{Complex configuration space: search speed and recall rate of different system configurations. The red line identifies the high-quality space where the configurations outperform the default setting. The stars mark the configurations that are optimal for both objectives.}
\label{fig:tradeoff_system}
\end{figure}

\begin{figure}
\centerline{\includegraphics[width=.45\textwidth]{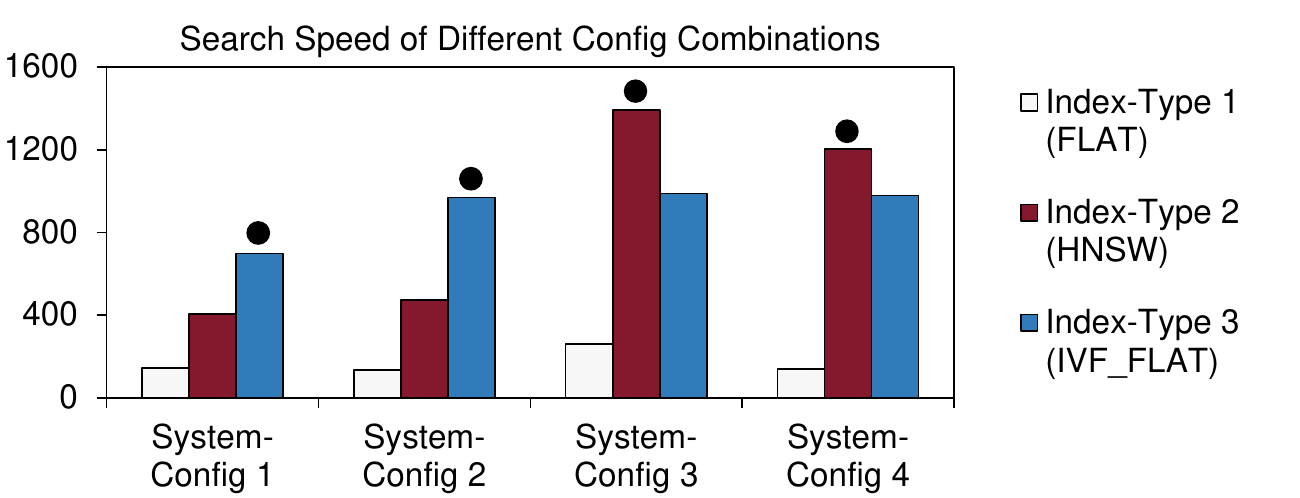}}
\caption{The best index type ($\bullet$) varies with system configs.}
\label{fig:inter_sys_index}
\end{figure}

Figure \ref{fig:tradeoff_system} shows an example of configuring two system parameters (\textit{segment\_maxSize} and \textit{segment\_sealProportion}). A darker color indicates a better search speed or recall rate. 
It can be seen that the performance of one parameter is impacted by another. For instance, most of the \textit{segment\_sealProportion} values (higher than 0.1) can lead to high search speed under a large \textit{segment\_maxSize} (= 1000), while the value needs to be higher than 0.9 if \textit{segment\_maxSize} is limited to 100. Similarly, the interdependence also exists between index and system parameters. As shown in Figure \ref{fig:inter_sys_index}, the best index type under different system configurations can be different: for system configuration 1 and 2, IVF\_FLAT is optimal, while HNSW becomes better under system configuration 3 and 4. This is because some index configurations may have higher segment size requirements in similarity search. 
Unfortunately, these interdependencies are not easy to understand even for experts, because VDMS is at a stage of rapid development. Milvus updates almost every few days or weeks, with frequent changes of parameters' number and ranges \cite{2023milvusgithub}. Therefore, a promising method should coordinately tune these parameters without requiring any prior domain knowledge.

\vspace{4pt}
\textbf{Challenge 2.} \textit{VDMS focuses on two important metrics: search speed and recall rate, and finding a configuration that strikes a good balance between the conflicting objectives is challenging.}
\vspace{4pt}

\begin{figure*}
    \centering
    \subfigure[Conflicting Objectives (Dataset 1)]{\includegraphics[height=0.2\textwidth]{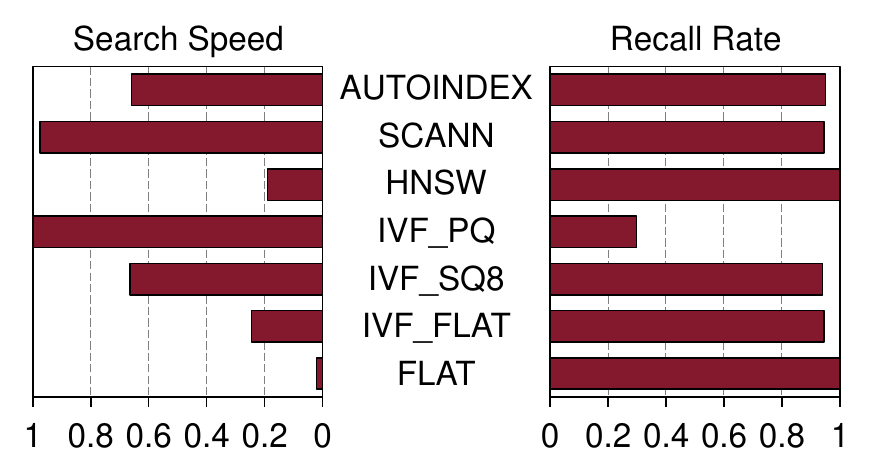}}
    \subfigure[Conflicting Objectives (Dataset 2)]{\includegraphics[height=0.2\textwidth]{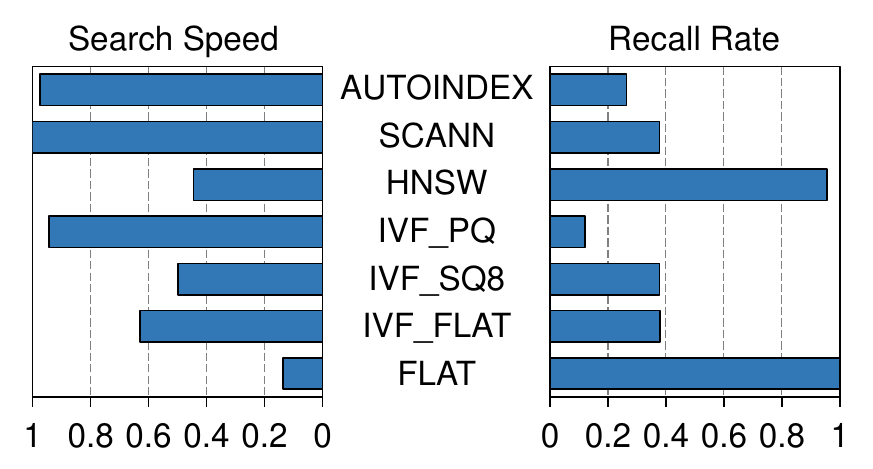}}
    \subfigure[Optimization Curves]{\includegraphics[height=0.2\textwidth]{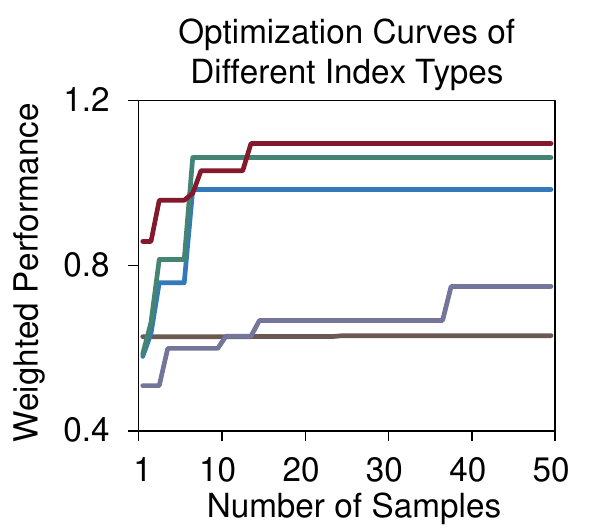}}
    \caption{ 
    \textbf{(a, b)} The best index type for search speed and recall rate can be very different (other parameters are fixed to their default values). 
    \textbf{(c)} Identifying the most suitable index type demands multiple tuning efforts.}
    \label{fig:interdependence}
    \vspace{-15pt}
\end{figure*}

As noted before, there are two important metrics (search speed and recall rate) to measure the performance of VDMS. However, it is difficult to find a configuration that is optimal for both metrics, since they are naturally conflicting. As presented in Figure \ref{fig:tradeoff_system}, the high-search-speed configurations and high-recall-rate configurations (marked by red lines) are very different from each other. Optimizing only one objective can lead to significant performance loss of the other. For example, the highest-recall-rate configuration has an unacceptable search speed, which is only 28.8\% of the default value. Therefore, an intelligent approach must strike a balance between the conflicting objectives, consequently finding optimal configurations without or less sacrificing any of the objectives (marked by red stars).

A naive approach to mitigate this difficulty is to simply fix the index type to the "generally recognized best one". Unfortunately, there are no winners in all scenarios \cite{li2019nowinners}. Figure \ref{fig:interdependence} (a) and (b) illustrates the performance of index types under two datasets (i.e., the stored data in VDMS for similarity search). If a user wants to maximize search speed while keeping recall rate higher than 0.8, then SCANN is a sufficiently good choice in dataset 1. However, in dataset 2, HNSW becomes the best one since most of the index types fail to keep reasonable recall rate.

\vspace{4pt}
\textbf{Challenge 3.} \textit{Tunable parameters are different for different index types, while identifying the most suitable index type within limited tuning budgets is challenging.}
\vspace{4pt}

There are many index types that can be selected in VDMS, however, the tunable parameters under each index type are different.
Table \ref{tab:index_types} shows the optional index types as well as the corresponding tunable parameters in Milvus. It can be seen that although some of the parameters are shared among different index types, the tunable parameters combinations are very different for different index types (e.g., IVF\_FLAT and IVF\_PQ both have parameter \textit{nlist} and \textit{nprobe}, while IVF\_PQ has unique parameter \textit{m} and \textit{nbits}). This could introduce extra complexity to auto-configuring VDMS, since most of the existing tuning methods \cite{van2017ottertune,cereda2021cgptuner,zhang2021restune,zhang2019cdbtune,li2019qtune} assume a fixed set of tunable parameters, that is, the parameters and their ranges do no change. 

\begin{table}
  \centering
  \caption{Index types and corresponding parameters in Milvus.}
  \resizebox{.48\textwidth}{!}{
    \begin{tabular}{|c|c|c|c}
    \hline
    \textbf{Supported} & \textbf{Description} & \textbf{Building \& Searching} \\
    \textbf{Index} &  & \textbf{Parameters} \\
    \hline
    FLAT  & Exhaustive approach & N/A ; N/A \\
    IVF\_FLAT & Quantization-based & nlist ; nprobe \\
    IVF\_SQ8 & Quantization-based & nlist ; nprobe \\
    IVF\_PQ & Quantization-based & nlist, m, nbits ; nprobe \\
    HNSW  & Graph-based & M, efConstruction ; ef \\
    SCANN & Quantization-based & nlist ; nprobe, reorder\_k \\
    AUTOINDEX & Default configuration & N/A ; N/A \\
    \hline
    \end{tabular}
    }
  \label{tab:index_types}%
\end{table}%

A natural idea to address this problem is to tune the parameters for each index type individually. However, it is time-consuming since we end up selecting only one index type disregarding the rest. In addition, optimizing only the optimal index type is also intractable, because it is difficult to find the best index type through simple sampling methods. Figure \ref{fig:interdependence} (c) shows the performance change of each index type with the number of samples (by uniform sampling). It can be observed that different index types have different degrees of performance variation, making it difficult to distinguish which one is optimal. For example, if we just select the best index type according to the first 10 samples, the green one would be the best. However, the red one is actually better as it surpasses the green one afterwards. Note that identifying the most suitable index type requires far more than 10 samples because we need to collect multiple samples for each index type before comparing them.


\subsection{Limitations of Existing Solutions}
So far, many auto-tuning solutions have been proposed to find a high-performance configuration for databases. However, owing to the challenges observed before, none of the existing solutions can solve the VDMS tuning problem very well.

\textbf{Heuristic Strategies.} Heuristic strategies include rule-based \cite{pgtune2023} and search-based \cite{ansel2014opentuner, zhu2017bestconfig} methods. Rule-based strategies design search rules relying on professional domain knowledge, which is difficult to obtain beforehand in VDMS tuning. Search-based strategies first sample a proportion of the whole configuration set, and then optimize around these sampled configurations. While search-based strategies incurs very low overhead even facing thousands of dimensions, they lack efficiency because the sampling process may miss better configurations within such a huge search space of VDMS tuning.


\textbf{Bayesian Optimization Strategies.} Bayesian optimization (BO) is an online learning-based method, which has been widely used in traditional database tuning \cite{van2017ottertune, cereda2021cgptuner, dalibard2017boat, zhang2021restune}. It employs a surrogate model to approximate the complex configuration-performance function and an acquisition function to recommend promising configurations. While BO can effectively speedup the search process towards optimal configuration, none of the existing BO-based database tuning solutions can strike the balance between search speed and recall rate in VDMS tuning. Moreover, they usually assume a fixed set of parameters, while in VDMS tuning, tunable parameters vary with the specified index type.

\textbf{Reinforcement Learning Strategies.} Reinforcement learning (RL) learns the tuning strategy by iteratively interacting with database system \cite{wang2021udo,zhang2019cdbtune,li2019qtune,ge2021watuning}. While RL-based solutions present excellent performance for high-dimensional \cite{zhang2019cdbtune}, query-aware \cite{li2019qtune} and workload-aware \cite{ge2021watuning} parameter tuning, they usually require very high tuning overhead or extensive offline training to obtain a well-behaved tuning agent. Moreover, existing RL-based database tuning solutions suffer from inefficiency and suboptimality when facing conflicting objectives and non-fixed action space in VDMS tuning.





\section{VDTuner: An Overview}

In this paper, we propose VDTuner, a framework adopting MOBO to automatically configure VDMS for maximizing both recall rate and search speed. In this section, we introduce the working principles of MOBO and summarize the advantages and challenges in designing an efficient MOBO based auto-tuner.

\subsection{Bayesian Optimization}



Bayesian optimization (BO) \cite{garnett2023bo} is a powerful sequential model-based optimization technique that aims to find the global optimum of an expensive, black-box function. The core idea behind BO is to construct a probabilistic \textit{surrogate model}, typically using Gaussian processes, to approximate the unknown objective function. This surrogate model is iteratively updated as new function evaluations are obtained, allowing for the incorporation of new information and the refinement of the model’s predictions.

In each iteration, BO uses an \textit{acquisition function}, such as expected improvement or probability of improvement, to determine the next point to evaluate. This acquisition function balances the exploration of unexplored regions and the exploitation of promising areas, enabling the algorithm to efficiently search for the global optimum with limited function evaluations.

\subsection{Multi-Objective Bayesian Optimization}

Multi-objective Bayesian optimization (MOBO) \cite{yang2019mobo, daulton2020qehvi} is an extension of BO, which addresses problems with multiple conflicting objectives. It aims to find a set of solutions that represents the Pareto front, which represents the optimal trade-offs between the different objectives. 

In MOBO, the surrogate model is extended to handle multiple objectives. This is straightforward by modeling and predicting each objective individually. The model captures the relationships between the input variables and the multiple objective functions, allowing for the prediction of objective values at unobserved points.

\begin{figure} 
\centerline{\includegraphics[width=.35\textwidth]{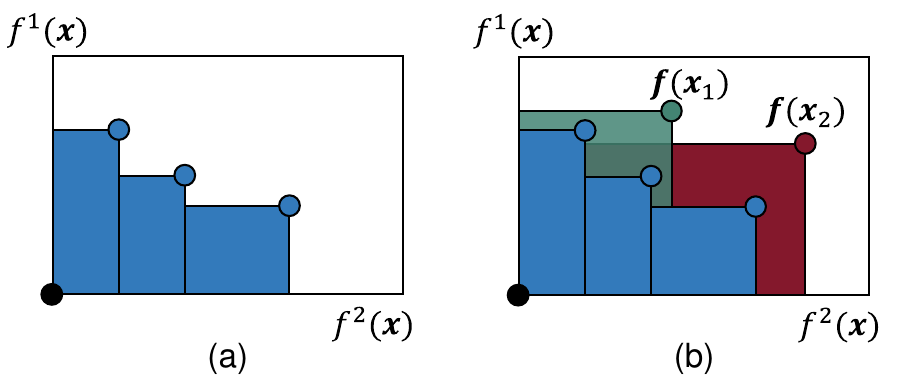}}
\caption{An illustration of EHVI. In (a), the blue area represents the hypervolume of three Pareto frontier solutions; in (b), the red area represents the EHVI of the newly added solution $\bm{x}_1$, and the green area represents the EHVI of the newly added solution $\bm{x}_2$; $\bm{x}_2$ has higher EHVI than $\bm{x}_1$, which will be considered as a better solution.}
\label{fig:HVI_illus}
\end{figure}

In MOBO, the acquisition function is extended to select promising solutions that strike a balance among the conflicting objectives. A popular approach is to use expected hypervolume improvement (EHVI) \cite{daulton2020qehvi}, which is a metric to assess the quality of a new solution by estimating the increase in hypervolume it would contribute when added to an existing set of solutions. To calculate EHVI, we first need to construct a hypervolume indicator, which quantifies the hypervolume of the current set of solutions. Then, the expected improvement in hypervolume is computed by integrating over the uncertain regions of the objective space and taking into account the probability distribution of the new solution. Among all the candidate solutions, the solution with highest EHVI is preferred by the acquisition function. Figure \ref{fig:HVI_illus} shows an example of how EHVI is computed.

The EHVI metric considers not only the improvement in individual objectives but also the overall coverage of the Pareto front, which represents the optimal trade-off solutions. By optimizing EHVI, we can guide the search towards solutions that improve the overall quality and diversity of Pareto front.

\begin{figure*}
\centerline{\includegraphics[width=.9\textwidth]{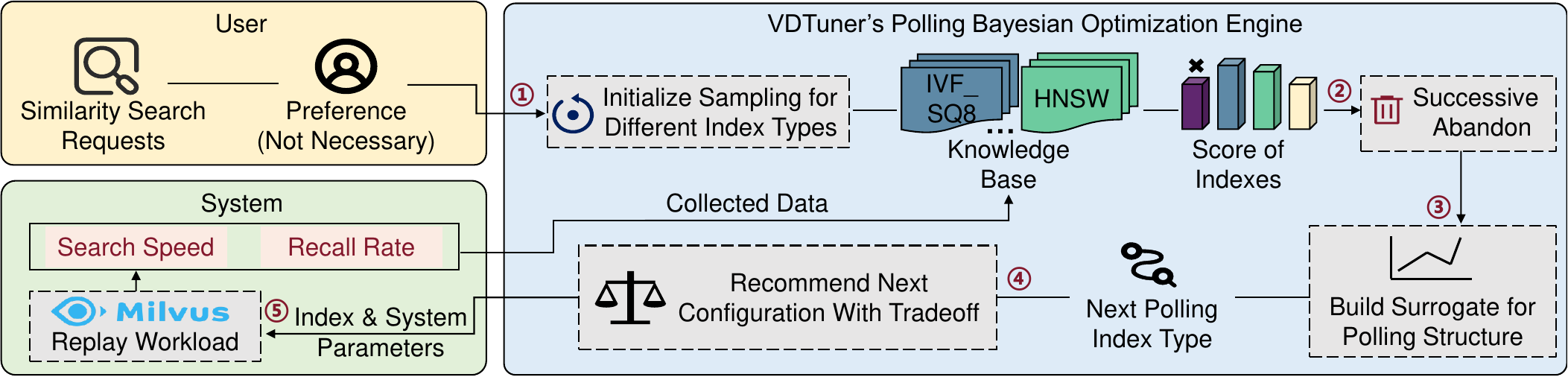}}
\caption{VDTuner's auto-configuration framework: holistically learning a Bayesian Optimization model, but selectively polling one index type at a time.}
\label{fig:architecture2}
\vspace{-15pt}
\end{figure*}


\subsection{Why MOBO is Suitable for VDTuner}

VDTuner adopts MOBO as its core optimization engine for the following primary reasons. 
First, MOBO does not require any prior knowledge about VDMS, which releases administrators’ pressure of understanding complex and rapidly changing VDMS versions.
Second, the evaluation of a VDMS configuration is expensive, often taking several minutes to even hours, especially when re-building the vector index after a change in index type. MOBO can circumvent the need for excessive configuration evaluations by exploring the complex multi-dimensional parameter space efficiently and intelligently. 
Third, MOBO was originally designed to optimize multiple objectives, which aligns perfectly with our need to optimize two objectives (search speed and recall rate).


\subsection{Challenges in Applying MOBO to VDMS Tuning}\label{sec:mobo_challenge}
Despite of MOBO’s many attractive advantages, applying it to VDMS tuning is still challenging for the following reasons. 
First, original BO model typically requires a fixed set of tuning parameters, while the tuning parameters are not fixed for different index types in VDMS. Therefore, a specialized design is needed in order to apply BO to VDMS. 
Second, our preliminary studies (Figure \ref{fig:interdependence} (c)) have shown that different index types have obvious performance differences. Therefore, allocating the tuning budgets to all index types equally is inefficient, and a more efficient budget allocating way is required. Third, there are some global tuning parameters (e.g., system parameters) shared by all index types in VDMS, implying that the knowledge learned from one index type may also be inspiring for other index types. Therefore, how to fully leverage the knowledge learned from different index types is worth exploring.

\section{VDTuner: Design and Implementation}\label{sec:vdtuner_desgin}

The overall workflow of VDTuner is described in Figure \ref{fig:architecture2}. Basically, VDTuner iteratively samples the configurations of VDMS to learn a holistic BO model which incorporates all tunable parameters of all index types. In each iteration, VDTuner specifies an index type and the acquisition function of the BO model recommends a configuration to be sampled for the specified index type. The index type is specified in a polling manner, but VDTuner successively abandons the index types with poor performance to ensure that more important index types receive a larger budget allocation. The newly sampled configuration is then used to update the surrogate model. The tuning process is terminated until a sufficiently good configuration is found. 

\subsection{Why Use a Holistic BO Model}

VDTuner aims to find the optimal configuration of VDMS, including the index type and the parameter setting for the index type. However, the tuning parameters associated with each index type are different. An intuitive approach is to build a separate BO model for each index type, and choose the best index type after all the BO models are well learned. However, this is not efficient, because first, many parameters (e.g., the system parameters) are shared among different index types, and the shared parameters will be tuned repeatedly; second, the knowledge about the shared parameters learned from different index types cannot be shared with each other.

In light of this, VDTuner incorporates the parameters of all index types in a holistic model (the shared parameters among different index types only have one copy), and tunes the parameters of different index types in a polling manner. This is more efficient because the shared parameters will not be tuned repeatedly. Moreover, since all index types share one BO model, the knowledge about the shared parameters learned from one index type is also useful for other index types. For example, the system parameter \textit{gracefulTime} is shared by all index types, which controls the bounded consistency level of VDMS. A small value of this parameter will lead to severe request blocking no matter which index type is chosen. By using a holistic BO model, VDTuner can learn this rule from previously sampled index types, which can greatly improve the tuning efficiency.

\subsection{Surrogate Model}
 
The surrogate model in BO serves as a proxy to approximate the unknown objective function based on the available data. In our context, the surrogate model is used to approximate the relationship between the tunable parameters and the performance of VDMS. A commonly used surrogate model is Gaussian process (GP), which is fully characterized by its mean function $m(\bm{x})$ and covariance function $k(\bm{x},\bm{x}')$, where $\bm{x}$ and $\bm{x}'$ represent input variables. Mathematically, a GP is defined as follows:
\begin{align}
    f(\bm{x}) \sim GP(m(\bm{x}), k(\bm{x},\bm{x}')).
\end{align}

Here, $f(\bm{x})$ represents a function drawn from the GP. The mean function $m(\bm{x})$ represents the expected value of $f(\bm{x})$ at any given input $\bm{x}$. The covariance function $k(\bm{x},\bm{x}')$ determines the similarity between function values at different inputs $\bm{x}$ and $\bm{x}'$. It captures the correlation structure of the function and controls the smoothness and variability of the GP. We choose Matérn 5/2 as the kernel function, owing to its excellent ability to balance the flexibility and smoothness when modeling the unknown function \cite{shahriari2015boloop}.

In our context, an input $\bm{x}$ refers to a configuration of the tuning parameters, including the index type, the index parameters of all index types, and the system parameters. Since we are concerned with two performance metrics, search speed and recall rate, VDTuner adopts a multi-output GP by assuming each output to be independent. Given an input $\bm{x}$, the GP model can estimate the performance (for both search speed and recall rate) of VDMS under this configuration. By updating the mean and covariance functions based on the continuously sampled configurations, the GP model will become increasingly accurate.




Despite of GP's theoretically-grounded prediction ability, roughly applying it to VDTuner is still inefficient. This is because at the early learning stage, the performance differences between different index types can vary significantly. The BO model is likely to exploit around the configurations recommended by the index types with good performance, rather than exploring the configurations recommended by the index types with bad performance. However, the index types with poor performance may not necessarily be bad because the current sampling is very limited, and it is not possible to judge the final performance accurately. This will increase the risk of getting trapped in local optimum.


To address this issue, VDTuner adopts a \textit{polling surrogate}, which normalizes the input data of GP by taking into account the performance variability of different index types. Instead of using the performance directly, we use a modified normalized performance improvement (NPI) \cite{cereda2021cgptuner} to evaluate the performance of configurations. Specifically, for a sampled configuration $\bm{x}_i$, let $(y^{spd}_i, y^{rec}_i)$ denote the performance of $\bm{x}_i$, where $y^{spd}_i$ denotes the search speed and $y^{rec}_i$ denotes the recall rate. The normalized performance of $\bm{x}_i$ is defined as:
\begin{align}
  (\hat{y}^{spd}_i, \hat{y}^{rec}_i) = (\frac{y^{spd}_i}{\overline{y}_t^{spd}}, \frac{y^{rec}_i}{\overline{y}_t^{rec}}), \label{eq:npi}
\end{align}
where $(\overline{y}_t^{spd}, \overline{y}_t^{rec})$ is a base performance value of index type $t$, which is set based on the most balanced non-dominated configuration achieved by index type $t$. 
Denote by ${\mathcal{X}}_t$ the set of sampled configurations of index type $t$, which are non-dominated (i.e., the performance of these configurations are not dominated by other sampled configurations). Let ${\mathcal{Y}}_t$ denote the set of performances of the configurations in ${\mathcal{X}}_t$,
then $(\overline{y}_t^{spd}, \overline{y}_t^{rec})$ is defined as:
\begin{align}
   (\overline{y}_t^{spd}, \overline{y}_t^{rec}) = \mathop{\arg\max}\limits_{(y^{spd},y^{rec}) \in \mathcal{Y}_t} \frac {1} {|y^{spd} / {y}_{max}^{spd} - y^{rec} / {y}_{max}^{rec}|}, \label{eq:2}
\end{align}
where ${y}_{max}^{spd}$ and ${y}_{max}^{rec}$ represent the maximum search speed and recall rate within ${\mathcal{Y}}_t$.

The \textit{polling surrogate} trains the GP model with the normalized performance, which reflects the relative improvement of a candidate configuration compared to the current best configuration for a specific index type. Therefore, it can eliminate performance differences between different index types and effectively prevent BO from getting trapped in local optimum.





\subsection{Acquisition Function}

In our context, the acquisition is used to recommend a configuration to be sampled for a specified index type. To this end, the acquisition function first sets the index type as the specified index type and sets the parameters not belonging to this index type as their default values, then recommends a configuration of the parameters belonging to this index type that achieves the maximum utility value according to the surrogate model's prediction. 


For a single-objective optimization, the acquisition function expected improvement (EI) is often applied to measure the utility values of configurations. Since we are concerned with two objectives, VDTuner employs the multi-objective generalization of EI, expected hypervolume improvement (EHVI) \cite{daulton2020qehvi,yang2019mobo} to recommend the configuration in VDMS tuning. 
Denote by ${\mathcal{Y}}$ the set of performances of all sampled non-dominated configurations, EHVI is defined as:
\begin{align}\label{eq:acq_ehvi}
    & \alpha_{EHVI}(\mathcal{X}', \bm{r}, \mathcal{Y}) \notag \\
    &= \mathbbm{E} \left[ HV(\bm{r}, \mathcal{Y} \cup \{\bm{f}(\mathcal{X}')\}) - HV(\bm{r}, \mathcal{Y}) \right] \notag \\
    &=  \int_{-\infty}^{\infty} (HV(\bm{r}, \mathcal{Y} \cup \{\bm{f}(\mathcal{X}')\}) - HV(\bm{r}, \mathcal{Y}))  \, d\bm{f},
\end{align}
where $\bm{r}$ is a two-dimensional (of search speed and recall rate) reference point; the $HV()$ function measures the hypervolume of the observed data; and $\bm{f}(\mathcal{X}')$ denotes the predicted performance values by the surrogate model. Thus, $\alpha_{EHVI}$ quantifies the expected hypervolume improvement of adding each candidate configuration within $\mathcal{X}'$.

We estimate Equation \ref{eq:acq_ehvi} using Monte Carlo Integration as the same of \cite{daulton2020qehvi}. 
We set $\bm{r} = 0.5 \cdot (\overline{y}_t^{spd}, \overline{y}_t^{rec})$ for each index type $t$, which implies that the objective value of an interested configuration should not be lower than the half value achieved by the most balanced configuration. This indicates to VDTuner that achieving a extremely high objective by sacrificing another one is not favoured. 

\subsection{Budget Allocation Among Index Types}
As discussed earlier, carefully allocating the tuning budgets to different index type is critical for VDTuner's efficiency. An easy yet effective way is to follow the round-robin rule: index types are assigned cyclically, ensuring each one takes turns in a sequential order, without favoring any one over the others. While round-robin is very suitable for situations where no prior information of index types are provided, it still lacks efficiency due to the inability to recognize a good index type. 

VDTuner performs a \textit{successive abandon} strategy to improve round-robin, where the index types are dynamically scored through a designed function and the worst one is successively abandoned during the tuning process. As VDTuner learns more about the configuration space, it gradually focuses its exploration on those promising index types. To achieve that effectively, the score function should fairly evaluate each index type's contribution to high performance sampling with careful tradeoff between two objectives. 

We construct a score function according to each index type's influence on HV of past observed performance. Specifically, if the calculated HV value is significantly reduced after excluding the data of a certain index type, it means that this index type contributes greatly to finding good configurations. 
For each index type $t = \{1,..,T\}$, with its observed performance $\mathcal{Y}_t$ of the non-dominated configurations, the HV influence of $t$ is calculated as the following subtracting form:
\begin{align}\label{eq:1}
    \Delta HV = HV(\bm{r}, \mathcal{Y}) - HV(\bm{r}, \mathcal{Y}/\mathcal{Y}_{t}),
\end{align}
where $\bm{r} = 0.5 \cdot \overline{\bm{y}}$, and $\overline{\bm{y}}$ is calculated same as Equation \ref{eq:2}, except that ${\mathcal{Y}}_t$ is replaced by the whole non-dominated configuration set ${\mathcal{Y}}$. Obviously, a higher $\Delta HV$ indicates a bigger contribution of index type $t$.In the implementation, since $HV(\bm{r}, \mathcal{Y})$ in Equation \ref{eq:1} is the same for all index types $t \in \{1,..,T\}$, we only need to calculate the scores as follows:
\begin{align}\label{eq:4}
    Score(t) = \max_{t' \in \{1,.,T\}}(HV(\bm{r}, \mathcal{Y}/\mathcal{Y}_{t'})) - HV(\bm{r}, \mathcal{Y}/\mathcal{Y}_{t}).
\end{align}

Deciding when to execute abandonment during the tuning process is also important to the performance. Giving up the index types too early may cause excellent index types to be discarded before they are well adjusted; while giving up too late may decrease the effect of budget allocation. VDTuner adopts a \textit{windowed variance} metric as the trigger condition of abandonment. Specifically, if the rank of an index type is consistently the worst (according to Equation \ref{eq:4}) lasting for a fixed-length window of iterations, it will be abandoned.

\subsection{Putting Them Together}

Overall, we report in Algorithm \ref{alg:1} the pseudo-code of VDTuner's polling Bayesian optimization. For a given workload (e.g., a batch of similarity search requests), VDTuner first performs an initial sampling for all index types (line 1-5), and the sampled configurations serve as the preliminary training data for VDTuner. In each tuning iteration (line 6-23), if there is more than one remaining index type, VDTuner first scores the index types and accordingly decides whether to abandon the worst index type (line 7-14); then, VDTuner builds a specialized GP surrogate model (line 15-18) using data from all index types; after that, for current polling index type, VDTuner recommends a promising configuration (line 19-21) that maximizes the acquisition function; finally, the recommended configuration is evaluated and VDTuner updates its knowledge base with the feedback (line 22). The termination condition of VDTuner is not fixed and can be flexibly specified, such as the maximum number of samples. 

\begin{algorithm}
\caption{VDTuner's polling Bayesian optimization.}
\label{alg:1}
\begin{algorithmic}[1]
\renewcommand{\algorithmiccomment}[1]{/* #1 */}

\renewcommand{\algorithmicrequire}{\textbf{Input}}
\REQUIRE Index type set $\{1,..,T\}$, index and system configuration space $\mathcal{X} \in \mathbbm{R}_d$, and workload for optimization.

\renewcommand{\algorithmicrequire}{\textbf{Initialize}}
\REQUIRE Observed data $\mathcal{D}_t = \emptyset$ for each index type $t$ and remaining index type set $\mathcal{T}_{remain} = \{1,..,T\}$.

\FOR{$t \in \{1,..,T\}$}
\STATE Initialize sampling for $t$ with its default configuration $\bm{x}_0$.
\STATE Replay the workload under $\bm{x}_0$.
\STATE Update $\mathcal{D}_t$ with $\bm{x}_0$ and observed performance $(y_0^{spd}, y_0^{rec})$.
\ENDFOR

\WHILE{True}
\IF{$len(\mathcal{T}_{remain}) > 1$} 
\FOR{$t \in \mathcal{T}_{remain}$}
\STATE Calculate $Score(t)$ according to Equation \ref{eq:4}.
\ENDFOR
\IF{Satisfy the windowed variance metric}
\STATE Remove $\mathop{\arg\min}_{t \in \mathcal{T}_{remain}} Score(t)$ from $\mathcal{T}_{remain}$
\ENDIF
\ENDIF
\FOR{$t \in \{1,..,T\}$}
\STATE Normalize $\mathcal{D}_t$ by Equation \ref{eq:npi}.
\ENDFOR
\STATE Build MOBO's surrogate with standardized data.
\STATE $t\_poll \Leftarrow $ next polling index type within $\mathcal{T}_{remain}$
\STATE Set search region $\mathcal{X}'$ for tunable parameters under $t\_poll$.
\STATE Generate next configuration $\bm{x}_{new} \in \mathcal{X}'$ by Equation \ref{eq:acq_ehvi}.
\STATE Reply the workload and update $\mathcal{D}_{t\_poll}$ with feedback.
\ENDWHILE

\renewcommand{\algorithmicrequire}{\textbf{Output}}
\REQUIRE Best found index type and configurations.

\end{algorithmic}
\end{algorithm}

\subsection{Handling User Preference}\label{sec:preference}
\textbf{Constraint Model.} So far, we have assumed that user has no preference on either objective of VDMS tuning. However, in some scenarios, users are likely to ask for optimizing search speed while keeping recall rate higher than a defined threshold, which can not be captured by EHVI acquisition function. Consequently, VDTuner incorporates a constraint model to guide the search within the recall rate constraint area. The constraint model quantifies the probability of candidate configurations satisfying the constraint. Specifically, when presented with a user-defined recall rate constraint (e.g., $rlim > 0.85$), we replace Equation \ref{eq:acq_ehvi} with a constraint EI acquisition function: 
\begin{align}
    &\alpha_{CEI}(\mathcal{X}_{cand}, rlim) = \alpha_{EI}(\mathcal{X}_{cand}) \cdot Pr(f^{rec}(\mathcal{X}_{cand})>rlim)  \notag\\
    &= \mathbbm{E} (\max (f^{spd}(\mathcal{X}_{cand}) - best\_f, 0)) \cdot Pr(f^{rec}(\mathcal{X}_{cand})>rlim) \label{eq:cei}.
\end{align}
The constraint acquisition function $\alpha_{CEI}$ is the product of an EI function, which measures the expected search speed improvement, and a probability function, which gauges the likelihood of achieving a recall rate higher than $rlim$. In addition, the base value $\overline{\bm{y}}$ in Equation \ref{eq:npi} is modified as the maximum function value achieved by index type $t$. This indicates VDTuner to relax the goal of achieving high search speed and recall rate simultaneously, instead focusing on maximizing search speed within constraint area.

\textbf{Bootstrapping with Previous Data.} For more general cases, users may have fluctuating recall rate preferences. Intuitively, learning from scratch for each new recall rate constraint is not efficient, since the previous sampled data may contain useful information that can be shared. In particular, the incipient samplings of the old recall rate constraint may reflect a rough performance distribution of the configuration space, even VDTuner gradually focuses on optimizing within constraint area afterwards. Hence, VDTuner bootstraps auto-tuning by warming up the surrogate model with previously sampled data (if available) of different recall rate constraints.

\section{Evaluation}
\subsection{Experiment Setting}
\textbf{Platform.}
We evaluate VDTuner on a popular VDMS, Milvus (version 2.3.1) \cite{wang2021milvus}, running on the platform shown in Table \ref{testbed}. Evaluated parameters for performance tuning are 16-dimensional in total, including index type, 8 index parameters (shown in Table \ref{tab:index_types}) and 7 system parameters (as recommended in Milvus documentation \footnote{https://milvus.io/docs/configure-docker.md\#purpose}).

\begin{table}
  \caption{Experimental Testbed.}
  \label{testbed}
  \centering
  \resizebox{.48\textwidth}{!}{
  \begin{tabular}{|c|c|}
    \hline
    \textbf{Component} & \textbf{Specification}\\
    \hline
    Processor & Intel(R) Xeon(R) Gold 5220 CPU  \\
    Processor Speed & 2.10GHz \\
    Logical Processor Cores & 72 Cores (36 physical cores) \\
    Private L1 \& L2 Cache Size & 32 KB and 1024 KB per core \\
    Shared L3 Cache Size & 24.75 MB \\
    Memory Capacity & 125 GB \\
    Operating System & CentOS 7.9.2009 with Linux 5.5.0 \\
   \hline
\end{tabular}}
\end{table}


\textbf{Workloads.}
We generate workloads using vector-db-benchmark \cite{2023vdbbench} and test three representative datasets shown in Table \ref{dataset}. The concurrent number of searching requests is set to 10 by default.
For each dataset, we send searching requests of top 100 similar vectors and calculate recall rate by comparing the result with correct results. The search speed is measured with throughput (i.e., request per second).

\begin{table}
  \caption{Evaluated Datasets.}
  \label{dataset}
  \centering
  \begin{tabular}{|c|c|c|c|}
    \hline
    \textbf{Dataset} & \textbf{Num. of Vectors} & \textbf{Dimension} & \textbf{Distance}\\
    \hline
    GloVe & 1,183,514 & 100 & Angular \\
    Keyword-match & 1,000,000 & 100 & Angular \\
    Geo-radius & 100,000 & 2048 & Angular \\
    \hline
\end{tabular}
\end{table}

\textbf{Baselines.} VDTuner is compared with a set of state-of-the-art auto-configuring methods. 
\begin{itemize}
    \item \textbf{Default} refers to make no effort on auto-configuring VDMS, instead applying default configuration in VDMS.
    \item \textbf{Random} \cite{bergstra2012randomsearch} sampling methods can be strong baselines for evaluating tuning algorithms due to their simplicity and effectiveness. Specifically, we adopt Latin hypercube sampling (LHS) \cite{loh1996latin} as a baseline approach, which is a space-filling method designed to uniformly distribute sample points throughout the value space.
    \item \textbf{OpenTuner} \cite{ansel2014opentuner} is a popular auto-configuring tool, which explores the configuration space using a pool of numerical approaches, collaborated by an AUC Bandit meta technique. To extend OpenTuner to tune VDMS scenario, we set the model's reward to the weighted sum performance of normalized search speed and recall rate.
    \item \textbf{OtterTune} \cite{van2017ottertune} adopts Gaussian Process Regression based optimization to auto-configure DBMS. Similarly, we use the weighted sum approach to optimize search speed and recall rate.
    \item \textbf{qEHVI} \cite{daulton2020qehvi} trades off between multiple objectives by measuring the expected hypervolume improvement of candidate configurations. The reference point of qEHVI is set to zero for each objective by default.
\end{itemize}

Since no prior works are dedicated to tuning unfixed parameters under different index types, we hypothetically assume the index type as a searching dimension to make the baselines suitable for optimizing multiple indexes simultaneously. The trigger condition of VDTuner's successive abandon is set as the occurrence of the worst-performing index type lasting for ten iterations. For OtterTune and qEHVI, we initialize the underlying BO models with 10 uniformly sampled configurations by LHS. The maximum time limit for workload replay is set to 15 minutes. For a failed configuration (that exceeds this time limit or causes VDMS to crash), we set the feedback values to the worst values in history to avoid the scaling problem \cite{van2021inquiry, zhang2022facilitating}. By default, we run 200 iterations for each method.

\subsection{Benefit of Auto-Configuration}
We first show the benefit of auto-configuring VDMS by VDTuner. We report the performance improvement of VDTuner compared with Default, which is defined as the maximum enhancement in search speed (or recall rate) without sacrificing recall rate (or search speed) relative to default performance. 
The results are presented in Table \ref{tab:benefit}. We have several observations.

\begin{table}
  \centering
  \caption{Performance improvement by auto-configuration.}
  \label{tab:benefit}
    \begin{tabular}{|c|ccc|}
    \hline
    &   &  Datasets & \\
    Metric  &	GloVe &	Keyword-match &	Geo-radius \\
    \hline
     Speed Improvement & 10.46\% & 11.17\% & 14.12\% \\
      Recall Improvement & 17.16\%	& 62.61\% &	186.38\% \\
    \hline
    \end{tabular}
\end{table}%

First, the default VDMS configuration has a considerable room for improvement, while VDTuner can significantly improve the performance, up to 14.12\% in search speed and 186.38\% in recall rate. Second, different datasets presents different improvement degrees (Geo-radius $>$ Keyword-match $>$ GloVe in our case). This is because the difficulty of similarity search depends on many factors such as data distribution and vector dimension. Geo-radius has an especially large vector dimension, where a good configuration (rather than the "commonly used" default setting) is very critical, hence, the auto-configuration shows the highest performance improvement. The results reconfirm the necessity of auto-configuring VDMS.

\begin{figure}
    \centering
    \subfigure[GloVe]{\includegraphics[width=0.475\textwidth]{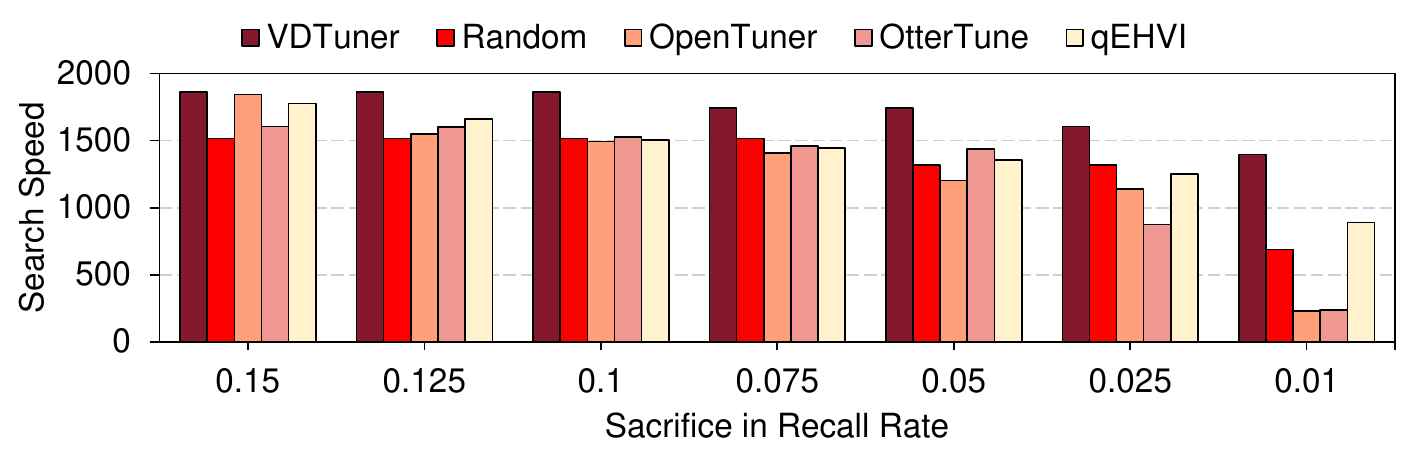}}
    \subfigure[Keyword-match]{\includegraphics[width=0.475\textwidth]{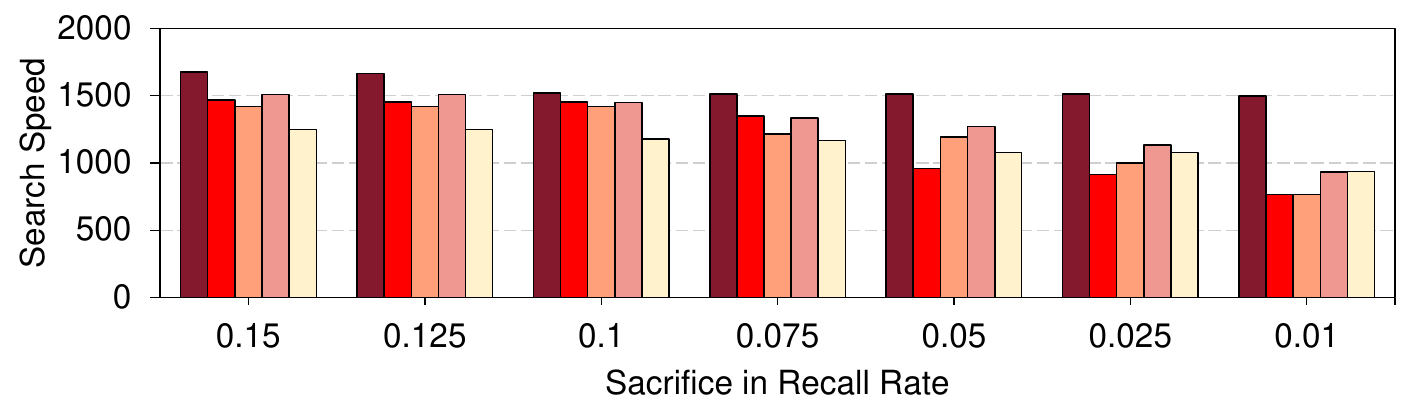}}
    \subfigure[Geo-radius]{\includegraphics[width=0.475\textwidth]{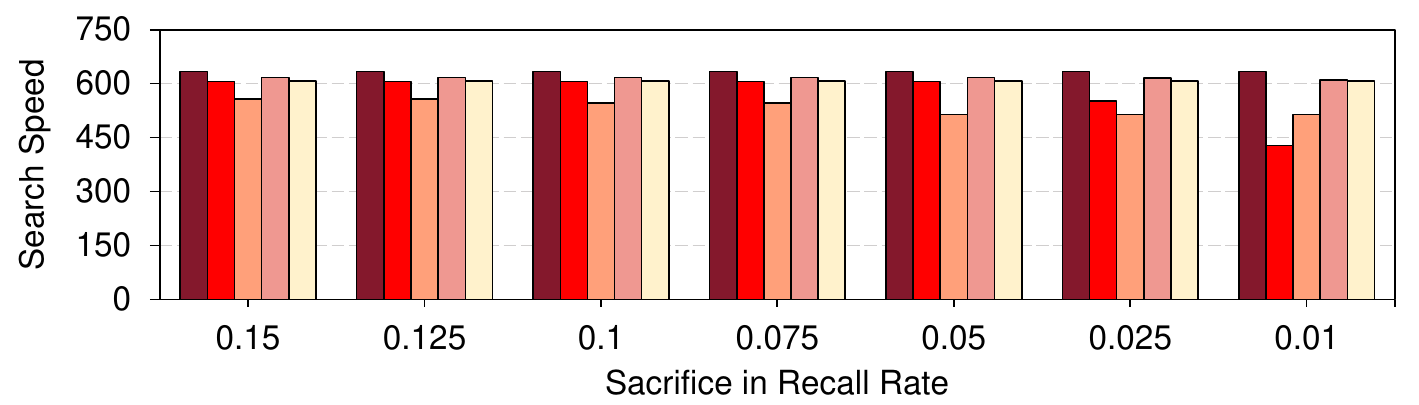}}
    \caption{VDTuner outperforms all competing baselines in terms of search speed and recall rate (on three datasets).}
    \label{fig:overall_sacrifice}
\end{figure}

\subsection{Tuning Efficiency}
\begin{figure*}
\centerline{\includegraphics[width=.98\textwidth]{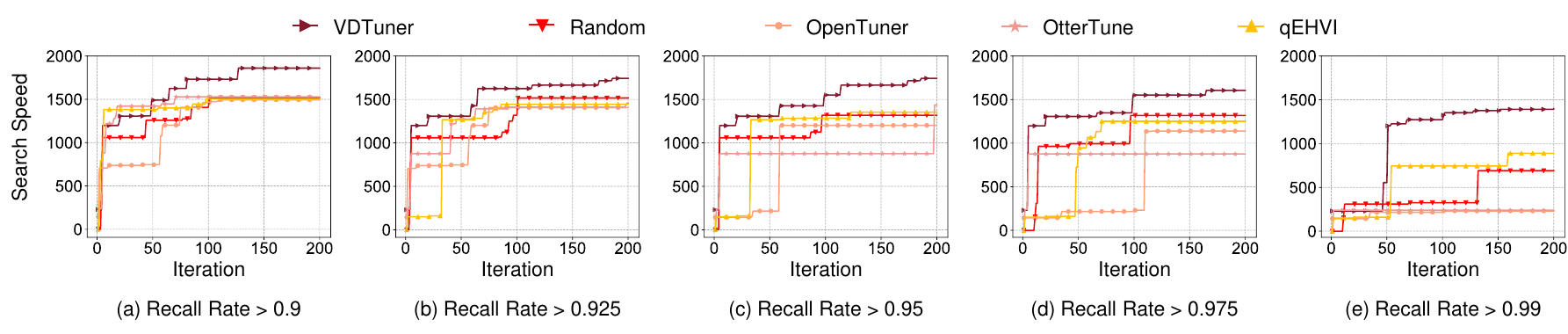}}
\caption{VDTuner achieves significantly better performance with a minimum number of samples compared to the baselines.}
\label{fig:exp_overall_speed}
\vspace{-15pt}
\end{figure*}

In this experiment, we compare the tuning efficiency of VDTuner and other competing baselines. Figure \ref{fig:overall_sacrifice} reports the best search speed achieved by different methods under different sacrifices in recall rate (from 0.85 to 0.99, with a step of 0.025). Note that we do not present the configurations of which recall rate is lower than 0.85, since a recall rate that is too low is often unacceptable in practical application scenarios. We have several observations.

\textbf{VDTuner strikes a balance between search speed and recall rates, surpassing the baselines in terms of both objectives.} We first note that VDTuner consistently achieves the best search speed under different levels of sacrificing recall rate. For instance, with recall rate sacrifice from 0.15 to 0.01 in Keyword-match, compared with the most competitive baseline, VDTuner achieves search speed improvement of 11.15\%, 10.35\%, 4.53\%, 12.19\%, 18.84\%, 33.32\% and 59.54\%, respectively. Overall, Random fails to find sufficiently good configurations since it can not utilize historical information. OpenTuner employs many numerical optimization techniques by assuming tuning parameters to be independent of each other, while the interdependency among VDMS parameters can result in a very uneven configuration space, where OpenTuner easily gets stuck into local optima. OtterTune employs a single-objective BO model to auto-tune the VDMS configurations, which can not provide reasonable tradeoff between conflicting search speed and recall rate. Despite qEHVI can strike the balance between objectives, it still lacks efficiency owing to the unawareness of the unstructured index configurations of VDMS.

Moreover, we find that VDTuner generally performs better within extremely difficult regions where recall rate is tightly limited. For all three datasets, with recall rate sacrifice tightening from 0.15 to 0.01, VDTuner's relative advantage against the most competitive baseline shows an upward trend, which is 1.03\% to 56.95\%, 11.15\% to 59.54\% and 2.64\% to 3.82\%, respectively. This indicates that VDTuner has superior ability to strike a balance between search speed and recall rate. For more quantitative analysis, we measure the tradeoff ability of all methods, which is defined as the standard deviation of search speed under different recall rate sacrifices. A lower deviation implies a better tradeoff between objectives. The order of tradeoff ability (best to worst) is VDTuner, qEHVI, OtterTune, OpenTuner and Random. This verifies that EHVI (used by VDTuner and qEHVI) plays an important role in trading off between search speed and recall rate. 





\textbf{VDTuner identifies better configurations markedly faster compared with competing baselines.} 
Figure \ref{fig:exp_overall_speed} displays the optimization curves of search speed by different methods with dataset GloVe. We observe that VDTuner finds sufficiently good configurations (i.e., configurations with performance better than the most competitive baseline) with the lowest number of samples and tuning time. Specifically, for recall rate sacrifice of 0.1, 0.075, 0.05, 0.025 and 0.01, VDTuner requires only 92\%, 64\%, 50\%, 69\%, 32\% of sampling number compared with baselines; while the advantage of tuning time is even bigger: 67\%, 47\%, 38\%, 49\%, 28\% (that is, up to 3.57× faster). The results proves that VDTuner not only presents the highest tuning efficiency, but also performs higher-quality sampling, where the configurations do no cause VDMS crashes or unreasonable index building time. 

\subsection{Why VDTuner Works Effectively}
In this section, we deep dive to analyze why VDTuner works so effectively. We first examine the main components in VDTuner, i.e., budget allocation and surrogate model, and then verify the effectiveness of VDTuner's holistic BO model.

\begin{figure}[htbp]
    \centering
    \subfigure[Different budget allocations]{\includegraphics[width=0.235\textwidth]{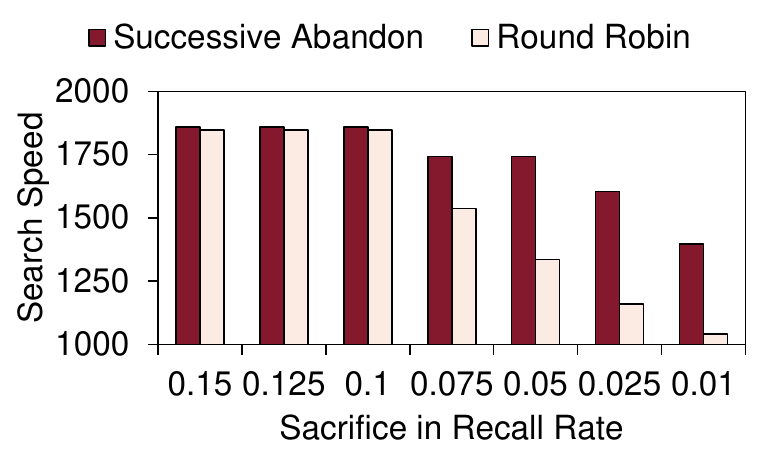}}
    \subfigure[Different surrogate models]{\includegraphics[width=0.235\textwidth]{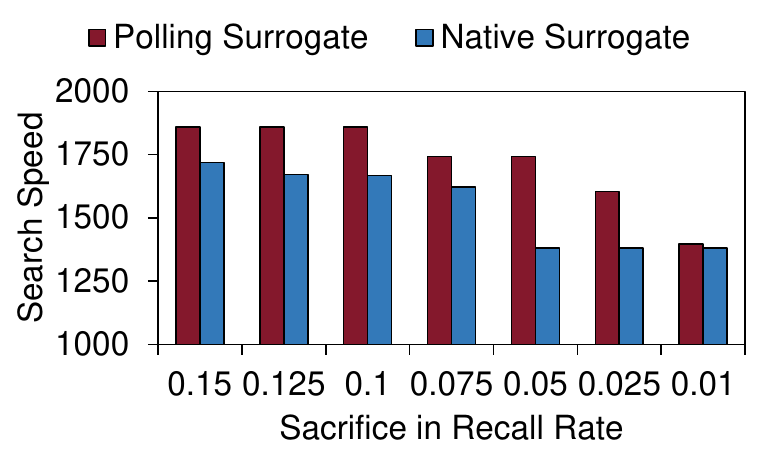}}
    \caption{Effectiveness of VDTuner's successive abandon and polling surrogate.}
    \label{fig:exp_ablation_sacrifice}
\end{figure}

\begin{figure}[htbp]
	\centering
	{\includegraphics[width=.46\textwidth]{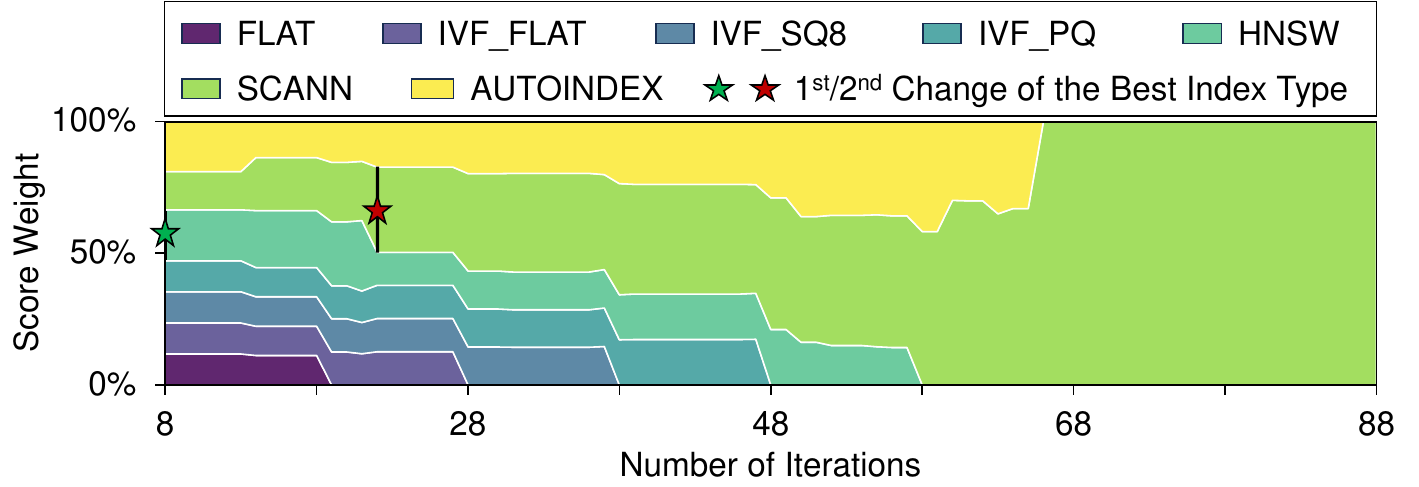}}
	\caption{VDTuner dynamically scores different index types during the tuning process. A weight of 0 represents this index type is abandoned. The star means the highest scoring index type has changed.}
	\label{fig:exp_ablation_hvi}
\end{figure}

\textbf{Effectiveness of Budget Allocation.}
Figure \ref{fig:exp_ablation_sacrifice} (a) reports the performance achieved by VDTuner's successive abandon strategy and its simplified version, round robin (in dataset GloVe). It can be seen that the successive abandon strategy brings search speed improvement under different recall rate sacrifices, which is up to 34\%. This can be attributed to VDTuner's ability of identifying the most suitable index type, which is measured by the dynamic score function. For a more detailed observation, Figure \ref{fig:exp_ablation_hvi} visualizes the dynamic scoring process, which depicts the weight of each index type as the number of samples increases. It can be observed that though HNSW has the highest score after the initial sampling, VDTuner gradually identifies the real best index type SCANN after learning more VDMS configuration information. 

%
%

\begin{figure}[htbp]
    \centering
    \subfigure[Native Surrogate]{\includegraphics[width=0.235\textwidth]{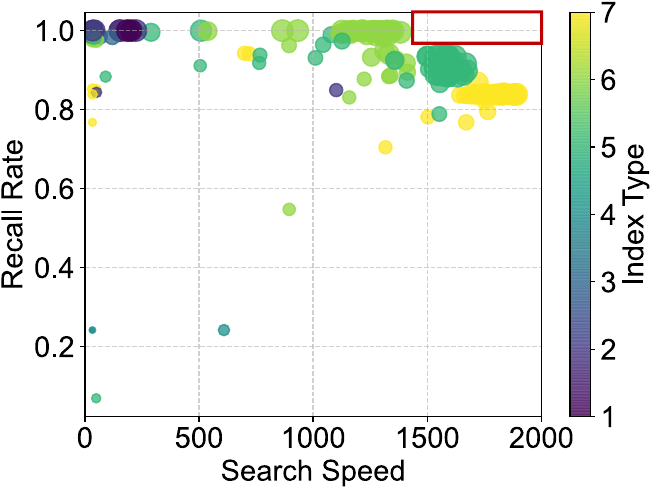}}
    \subfigure[Polling Surrogate]{\includegraphics[width=0.235\textwidth]{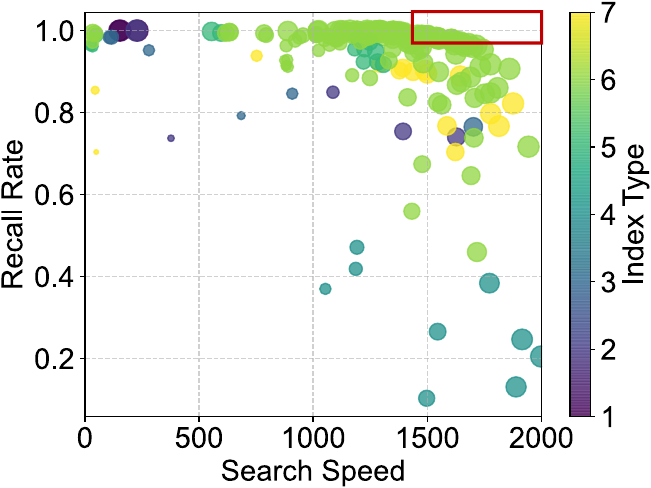}}
    \caption{VDTuner's polling surrogate results in wider exploration and higher quality sampling. The color and size of the configurations identify the index type and Pareto rank (the higher the rank, the larger the size), respectively. Red rectangles indicate both high search speed and recall rate.}
    \label{fig:exp_ablation_surrogate_scatter}
\end{figure}

\textbf{Effectiveness of Surrogate Model.}
We next study the effectiveness of VDTuner's polling surrogate model. Figure \ref{fig:exp_ablation_sacrifice} (b) shows the performance improvement of polling surrogate over a native Gaussian process surrogate. We observe that the polling surrogate brings obviously better search speed (up to 26\% improvement) under different recall rate sacrifices. For a more detailed observation, we depict in Figure \ref{fig:exp_ablation_surrogate_scatter} all of the configurations sampled by two methods. We have several observations. First, the two surrogates both choose more SCANN, AUTOINDEX and HNSW as their index types. Second, the polling surrogate explores wider space of different recall rate values, while the native surrogate presents more similar performance sampling (showing a cluster shape for each index type). Third, as a result, the polling surrogate guides a higher-quality search where both search speed and recall rate are high (marked by red rectangles). The results imply that VDTuner effectively learns multiple index types jointly, thus achieving better performance.

\textbf{Holistic BO Model VS. Optimizing Each Index Type Individually.} We then compare VDTuner with the approach which tunes the parameters for each index type individually. The results show that in our selected experimental scenario, the index type selected by VDTuner is the same with that selected by the approach tuning the parameters for each index type individually, and the parameters generated by these two approaches are very close. For example, they both select SCANN as the best index type, and for over 80\% of the parameters, the difference between the parameters generated by the two approaches is less than 5\% (the difference is defined as the absolute difference divided by the scale of that parameter). Note that the parameters generated by the two approaches don't necessarily have to be identical, as there isn't just one optimal parameter configuration.

In order to understand the changes of index selection and parameters in different datasets, we summarize in Table \ref{tab:param_change} the index type and representative parameters recommended by VDTuner for different datasets. We observe that the selected index type varies for different datasets (SCANN is selected for GloVe and Keyword-match, HNSW is selected for ArXiv-titles). This is because the data characteristics (such as dimensionality, distribution, and density) would influence the efficiency and effectiveness of different index types, making certain index types more suitable for specific datasets than others. We also observe that the parameters generated by VDTuner vary greatly across different datasets, even for the datasets with the same index type (e.g., GloVe and Keyword-match). The reason is that different data characteristics require different parameter settings to achieve the optimal performance. For example, vector search is more challenging for the datasets with lower correlation between dimensions (e.g., Keyword-match), requiring a higher nprobe (the parameter controlling  the number of candidate cluster centers used in each query for searching) to enlarge the search scope in order to achieve a high recall rate. The results confirm the necessity of VDMS performance tuning for different workloads.

\begin{table}[htbp]
	\centering
	\caption{Changes of index and parameters across different datasets.}
	\resizebox{.48\textwidth}{!}{
		\begin{tabular}{|c|ccc|}
			\hline
			&  & \textbf{Datasets} & \\
			& {GloVe} & {ArXiv-titles} & {Keyword-match} \\
			\hline
			Index Type  & {Index: SCANN} & {Index: HNSW} & {Index: SCANN} \\
			and Parameters & {nlist: 301} & {M: 64} & {nlist: 680} \\
			of the Best    & {nprobe: 36} & {efConstruction: 194} & {nprobe: 238} \\
			Configuration & {reorder\_k: 283} & {ef: 100} & {reorder\_k: 465} \\
			\hline
		\end{tabular}%
	}
	\label{tab:param_change}%
\end{table}%

\begin{figure}[htbp]
	\centering
	\includegraphics[width=.48\textwidth]{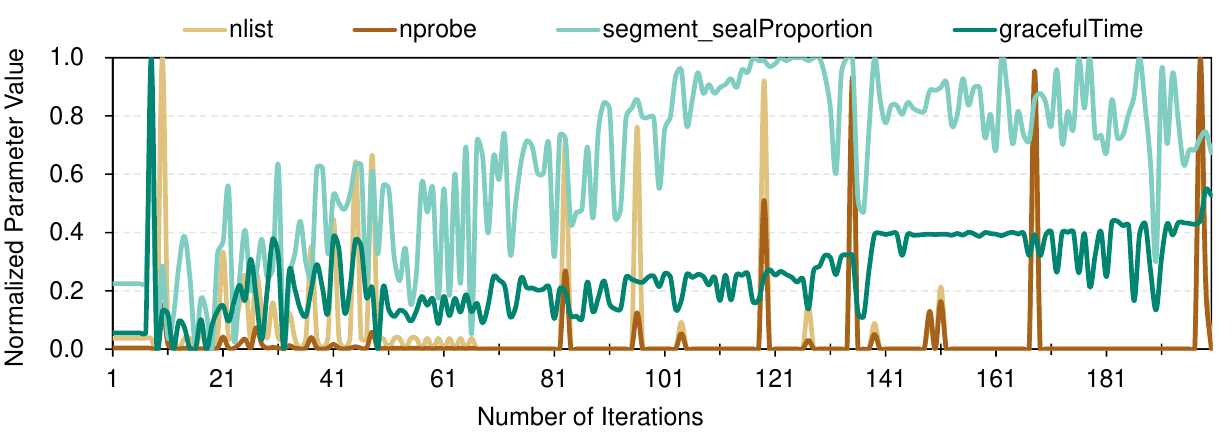}
	\caption{Changes of parameters with the number of iterations.}
	\label{fig:parameter_change}
\end{figure}

We have also recorded the changes of parameters generated by VDTuner with the number of iterations (for dataset Geo-radius). Figure \ref{fig:parameter_change} shows the results. As can be seen, in the early stage of the tuning process, the fluctuations of all parameters are relatively large. As the number of iterations increases, all parameters basically converge to fluctuate within a small range. This aligns with the characteristics of Bayesian Optimization, where early stages tend to favor exploration while later stages lean towards exploitation. Note that there is also large fluctuations occasionally in the late stage, which is reasonable because Bayesian Optimization will continue to explore persistently, despite the probability of exploration becomes increasingly smaller. 

\subsection{Scalability}
\textbf{Larger Datasets.} We next validate the effectiveness of VDTuner on a much larger dataset (deep-image, 10x bigger than GloVe). We compare the performance of VDTuner with the top-performing baseline (i.e., qEHVI) in our selected scenario. The results indicate that VDTuner still maintains significant advantages in performance. Specifically, when the sacrifice in recall rate is 0.99, VDTuner achieves a performance improvement of 159\% in search speed, resulting in an $8.1\times$ faster tuning speed when reaching the same level of performance.

\begin{figure}
\centerline{\includegraphics[width=.45\textwidth]{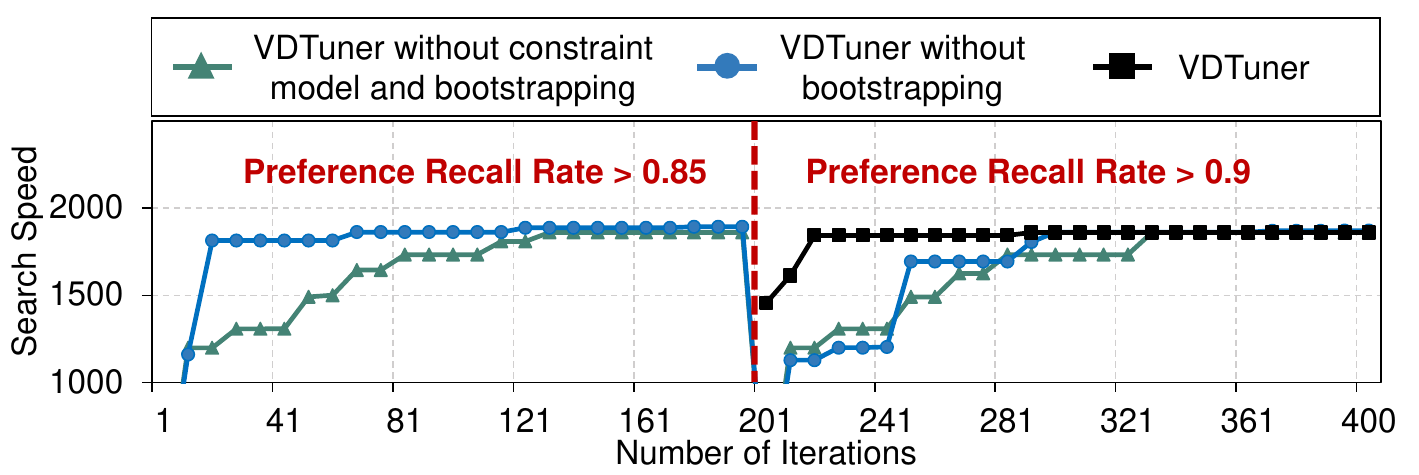}}
\caption{VDTuner flexibly handles user preference on different recall rates.}
\label{fig:exp_case_preference}
\end{figure}

\textbf{Handling User Preference.}
We next verify the effectiveness of VDTuner when users have specific preference on recall rate. Three versions are considered: (1) VDTuner without constraint model and bootstrapping, which means the user preference of recall rate is not modeled (instead directly optimizing both objectives) and historical data (i.e., data of optimizing previously appeared recall rate) is not used; (2) VDTuner without bootstrapping, which means historical data is not used; (3) VDTuner, which is the complete version. We consider optimizing scenarios with preference recall rates $>$ 0.85 and $>$ 0.9 in sequence, with each for 200 iterations. The results are shown in Figure \ref{fig:exp_case_preference}.

First, we observe that the constraint model brings significantly better tuning efficiency. For recall rate $>$ 0.85 and $>$ 0.9, VDTuner with constraint model requires only 49\% and 75\% of the samples (1.87$\times$ and 1.30$\times$ faster in terms of tuning time), respectively, to achieve the same performance as VDTuner without constraint model. This is because VDTuner with constraint model can focus more on optimizing search speed as long as it finds recall rate higher than the threshold, while ignoring this pre-defined threshold may lead to broader exploration for different recall rate levels and is therefore slower. We also note that the advantage of constraint model is relatively greater for a looser recall rate constraint (recall rate $>$ 0.85). This is because a tighter constraint leads to reduced feasible region and increased search difficulty, thus requiring more tuning efforts even if VDTuner models the constraint. 

Moreover, it can be seen that the bootstrapping technique further improves the auto-configuration efficiency on the basis of VDTuner with constraint model. The complete version of VDTuner requires only 66\% of the samples (while the value is 75\% if not using bootstrapping) compared with VDTuner without constraint model and bootstrapping for recall rate $>$ 0.9. This is because VDTuner bootstraps the auto-configuration by warming up the surrogate model using the historical data of optimizing recall rate $>$ 0.85, which provides high-quality initial configurations and an approximate exploration space distribution.

\textbf{Cost-Effectiveness Optimization.}
In some practical scenarios, users may be less in need of extremely optimized search speed. Instead, they are more concerned with cost-effectiveness, which requires extra consideration of the memory usage in VDMS. To this end, we study a cost-effectiveness optimization case, where the objective of search speed (QPS) is replaced by cost-effectiveness (QP\$). We simply assume a linear charges in memory usage with \$$\eta$ per second$\cdot$GiB, then QP\$ is defined as:
\begin{align}
    \text{Cost-Eff.} = \frac{\text{Search Speed (query/sec)}}{\text{Price (\$/sec)}} = \frac{\text{Search Speed}}{\eta \cdot \text{Memory Usage}}.
\end{align}
Since VDTuner trains the model with normalized function values, the value of $\eta$ will not affect the results. In the implementation, we set $\eta = 1$. Note that optimizing other resources and price functions is trivial by modifying the definition of Cost-Eff. Our work is not limited by any specific resource or price function.

\begin{figure}
    \centering
    \subfigure[Performance of optimizing cost effectiveness (relative to optimizing search speed).]{\includegraphics[width=0.475\textwidth]{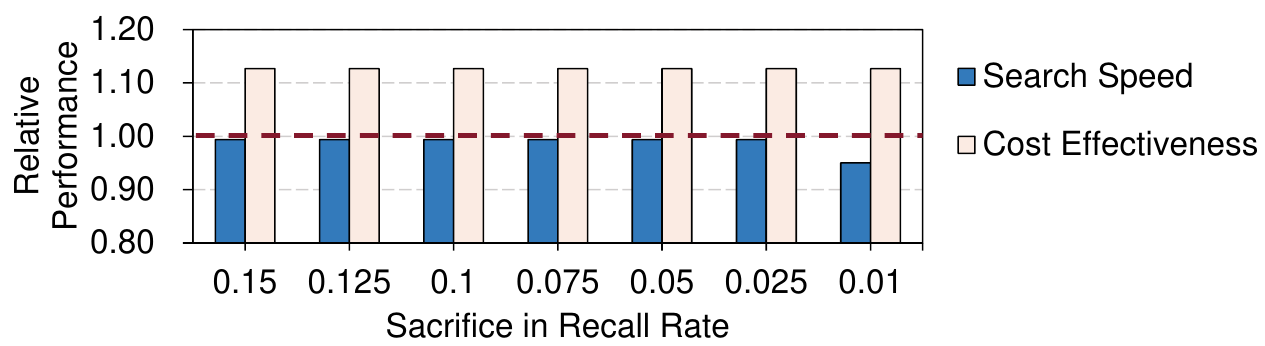}}
    \subfigure[Parameter's contribution to memory usage and search speed.]{\includegraphics[width=0.475\textwidth]{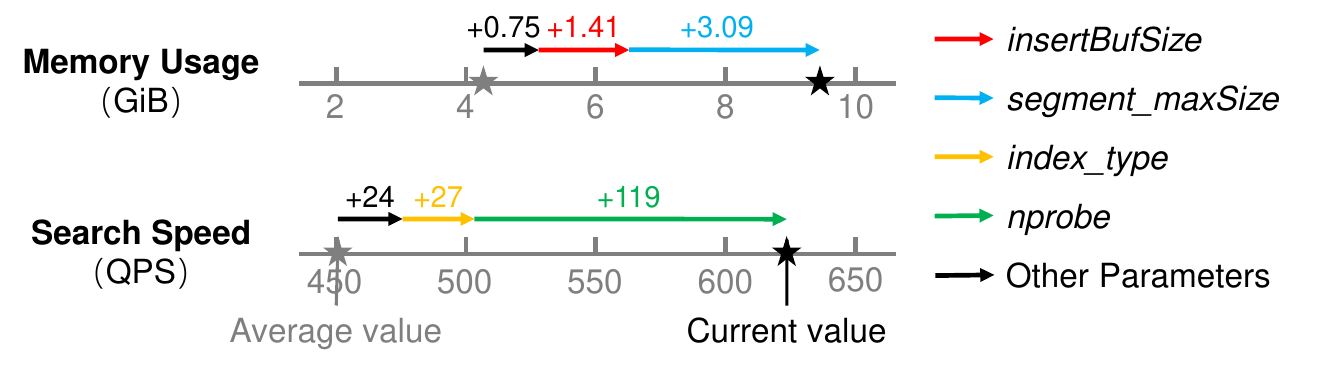}}
    \caption{Comparison between optimizing cost effectiveness and search speed.}
    \label{fig:exp_mem}
\end{figure}

To compare the two optimization objectives, we record the performance and memory usage of optimizing QP\$ and QPS, respectively (for dataset Geo-radius). As shown in Figure \ref{fig:exp_mem} (a), we observe that optimizing QP\$ results in obviously higher QP\$ (up to 13\%) and and lower QPS (up to 5\%), proving that QP\$ is perceptible to memory usage. We further compare the memory usage of all configurations sampled by optimizing QP\$ and QPS, respectively. As expected, optimizing QP\$ leads to markedly lower memory usage than optimizing QPS (3.89 GiB $\pm$ 1.75 VS. 5.19 GiB $\pm$ 2.44). 

To deep dive, we use a game theory method, SHAP path \cite{lundberg2017shap}, to evaluate each parameter's influence on performance and memory usage, respectively. As shown in Figure \ref{fig:exp_mem} (b), the most important parameters for memory usage and search speed are \textit{segment\_maxSize} (+3.09 GiB) and \textit{index\_type} (+119 QPS), respectively. The search-speed-optimization version recommends a large \textit{segment\_maxSize} for higher search speed no matter how high the memory usage it will cause. In contrast, the cost-effectiveness-optimization version makes a tradeoff between search speed and memory usage, and recommends a relatively small \textit{segment\_maxSize}, which significantly reduces memory usage while sacrificing a little opportunity to explore high QPS. The results demonstrate that VDTuner is scalable to cost-aware optimization objectives.


\subsection{Overhead}
We next evaluate the overhead of VDTuner's multi-objective Bayesian optimization engine. 
We record the breakdown of tuning time (including target workload replay and configuration recommendation) of each method for 200 iterations in dataset GloVe. Results are reported in Table \ref{tab:overhead}. 
We have several observations.

First, Random has the longest total tuning time, implying that it lacks the ability to find high-quality configurations (usually with reasonable data loading/index building time). OpenTuner takes slightly shorter total tuning time than VDTuner because it samples more sub-optimal index types that have poor performance despite of slightly shorter index building time. Second, the configuration recommendation time of VDTuner is similar to qEHVI and mildly more than OtterTune, attributing to VDTuner's components: budget allocation and multi-objective acquisition function, respectively. Third, the configuration recommendation time of all methods only accounts for a small fraction of the total tuning time (e.g., 1.44\% for VDTuner), which is acceptable given the superior performance of VDTuner.


\begin{table}
  \centering
  \caption{Time breakdown for 200 iterations of each method.}
    \begin{tabular}{|c|c|c|c|}
    \hline
           & Configuration & Workload &  \\
    Method & Recommendation &  Replay & Total \\
    \hline
    VDTuner & 438s (1.44\%) & 30,034s & 30,472s \\
    Random & 2s (0.00\%) & 42,860s & 42,862s \\
    OtterTune & 173s (0.41\%) & 42,045s & 42,218s \\
    qEHVI & 406s (1.02\%) & 39,381s & 39,787s \\
    OpenTuner & 7s (0.03\%) & 27,632s & 27,639s \\
    \hline
    \end{tabular}%
  \label{tab:overhead}%
\end{table}%


\section{Related Work}
\textbf{Vector Data Management System.} 
Many libraries \cite{johnson2019faiss,malkov2018hnswlib,annoy2023} and vector data management systems \cite{chroma2023,qdrant2023,wang2021milvus,guo2022manu} have been developed for similarity search. 
Milvus \cite{wang2021milvus} is a widely used, purpose-built VDMS which supports billion-scale data, dynamic update and heterogeneous computing platform. Later, Manu \cite{guo2022manu} provides a cloud native VDMS solution to achieve elasticity. 
While most of the VDMS rely on the users to specify the index and system configurations, Manu provides an auto-tuning component of index parameters. 
However, it only supports the auto-configuration under a specific index type and requires a user-defined utility function over objectives. 
In contrast, VDTuner simultaneously optimizes many index types together as well as system configurations, and automatically strikes the balance between search speed and recall rate.

\textbf{Auto-Configuration of Database.} Automated configuration of traditional databases \cite{van2017ottertune, zhang2019cdbtune, trummer2022dbbert, zhang2023unitune, pgtune2023, li2019qtune, cereda2021cgptuner} or big data analytical systems \cite{zhao2022dremel,xin2022locat,lin2022icdetuning,herodotou2022icdetuning} has proven to have great potential in improving database performance. The common tuning methods include heuristics \cite{pgtune2023, ansel2014opentuner, zhu2017bestconfig}, Bayesian optimization \cite{van2017ottertune, cereda2021cgptuner, dalibard2017boat, zhang2021restune}, and reinforcement learning \cite{wang2021udo,zhang2019cdbtune,li2019qtune,ge2021watuning}. 
Unfortunately, these methods can not be directly used by VDTuner since VDMS tuning requires extra efforts in trading off between search speed and recall rate and optimizing different index types together.



\section{Conclusion}

In this paper, we propose VDTuner, a learning-based performance tuning framework that optimizes VDMS index and system configurations. VDTuner actively strikes the balance between search speed and recall rate, and delivers better performance via a polling structure, a specialized surrogate model and an automatic budget allocation strategy. Extensive evaluations prove that VDTuner is effective, beating the baselines by a significant margin in terms of tuning efficiency, as well as scalable for fluctuating user preferences and cost-aware objectives. In the future, we would like to extend VDTuner to an online version to actively capture different workloads. Moreover, we also want to extend it to optimize more levels (e.g., data partition) of VDMS to further improve the performance and resource utilization.




 \section*{Acknowledgment}
This work was supported in part by the National Science Foundation (NSF) of China (grant numbers 62293510/62293513, 62272252, 62272253, 62141412); in part by the NSF of Tianjin 21JCYBJC00070; and in part by Ant Group through CCF-Ant Research Fund.

\bibliographystyle{IEEEtran}
\bibliography{IEEEabrv,ref.bib}

\vspace{12pt}

\end{document}